\documentclass[aps,prd,reprint,superscriptaddress,preprintnumbers,longbibliography]{revtex4-1}

\usepackage{amsmath,amssymb,bm,mathrsfs,bbold,dsfont}
\usepackage{xfrac}
\usepackage[dvipsnames]{xcolor}
\usepackage{slashed}
\usepackage{multirow}
\usepackage{booktabs}

\usepackage{epsfig}
\usepackage{psfrag}
\usepackage{tikz}
\usepackage{tikz-feynman}
\usepackage[caption=false]{subfig}  
\usetikzlibrary{arrows.meta,positioning,calc,backgrounds, shapes.arrows}

\PassOptionsToPackage{caption=false}{subfig}

\usepackage[sort&compress]{natbib}
\usepackage[colorlinks=true,
    linkcolor=red,
    citecolor=blue,
    urlcolor=blue,
    filecolor=blue,
    anchorcolor=blue,
    menucolor=blue,
    linktocpage=true,
    pdfproducer=medialab,
    pdfa=true
]{hyperref}
\usepackage{cleveref}

\usepackage{setspace}
\usepackage{relsize}
\usepackage{enumerate}
\usepackage{comment}
\usepackage[normalem]{ulem}
\usepackage{srcltx}
\usepackage{xcolor}
\usepackage{diagbox}

\definecolor{mypurple}{HTML}{5D6DC1}
\definecolor{myorange}{HTML}{f19700}
\definecolor{myred}{HTML}{e65c0d}
\definecolor{navyblue}{HTML}{0055CC}

\newcommand{\Fig}[1]{Fig.~\ref{#1}}
\newcommand{\Eq}[1]{Eq.~\eqref{#1}}

\newcommand{\be}{\begin{equation}}
\newcommand{\ee}{\end{equation}}

\newcommand{\ga}{g_{a\gamma \gamma}}

\newcommand{\TRH}{T_{\rm RH}}
\newcommand{\tRH}{t_{\rm RH}}
\newcommand{\mKK}{\Delta m_{\rm KK}}
\newcommand{\GeV}{{\rm GeV}}

\newcommand{\keV}{{\rm keV}}

\newcommand{\meV}{{\rm meV}}

\newcommand{\emphb}[1]{\textbf{\emph{#1}}}

\begin{document}

\preprint{CERN-TH-2026-169, MIT-CTP/6073}

\title{The Dark Dimension meets the Axiverse}

\author{Kevin Langhoff}
\affiliation{Massachusetts Institute of Technology (MIT), Cambridge, Massachusetts 02139, USA}

\author{Maria Ramos}
\affiliation{CERN, Theoretical Physics Department, Esplanade des Particules 1, Geneva 1211, Switzerland}

\author{Mario Reig}
\affiliation{CERN, Theoretical Physics Department, Esplanade des Particules 1, Geneva 1211, Switzerland}

\begin{abstract}\noindent
We explore the cosmological implications of combining {dark dimension} scenarios with an axiverse. If gauge sectors are realized on branes, \emph{towers} of Kaluza-Klein (KK) excitations of closed string axions can propagate through the dark dimension in addition to the tower of graviton excitations. This modifies cosmology in two ways. First, if any of these axion towers interact with the standard model (SM) plasma, they can significantly alter the freeze-in production of the cosmological abundance of tower states. Freeze-in to graviton and axion towers can provide all of dark matter (DM) for an axion decay constant in $10^{12} \text{ GeV }\lesssim f_a\lesssim 10^{16}$ GeV and reheating temperatures $5\text{ MeV }\lesssim \TRH \lesssim O(1)$~GeV. Second, different towers \emph{fragment} into each other and redistribute energy; each tower's fraction of energy at late times is fixed by their interactions. If there are $N\gg1$ axion towers, the energy visibly injected into the SM by any decaying tower is diluted by a factor of $N$. This suppression offers a simple realization of how dark dimension dark matter can avoid strong cosmological constraints which rule out the simplest models. In the process of this exploration we develop a continuum approach to evaluating tower fragmentation which offers insight and aids numerical calculations by reducing the problem to quadrature.

\end{abstract}

\maketitle

\section{Introduction}

\emph{Swampland} principles \cite{Ooguri:2006in,Lee:2019wij,Lee:2019xtm,Lust:2019zwm,Obied:2018sgi,Bedroya:2019snp} together with the observed cosmological constant, $\Lambda^{1/4}\approx 2.3~{\rm meV}$, hint at the existence of a single micron-scale extra dimension \cite{Montero:2022prj,Vafa:2024fpx}. This \emph{Dark Dimension} scenario predicts a tower of Kaluza-Klein (KK) graviton excitations with uniform mass splitting $\Delta m_{\rm KK}\sim \Lambda^{1/4}$, which can source the dark matter (DM) of the universe \cite{Gonzalo:2022jac}. The tower of KK gravitons is unavoidably produced via UV freeze-in through Planck-suppressed interactions \cite{Giudice:1998ck, Hall:1999mk}
\be\label{eq:KK_graviton_coupling}
    \mathcal{L} \supset -\frac{1}{M_P}h^{(n)}_{\mu \nu} T^{\mu \nu}\,,
\ee
where $M_P$ is the 4D reduced Planck mass 
and $n$ stands for the $n^{\rm th}$ mode of the KK tower.

The freeze-in abundance is set by the temperature $\TRH$ just after reheating. The observed DM abundance follows for $\TRH \sim \mathcal{O}({1\,\rm GeV})$~\cite{Gonzalo:2022jac}, a value kept low by the enormous number of accessible modes, $\TRH/\Lambda^{1/4}\sim \mathcal{O}(10^{12})$. At production, the energy density of the gravitons is dominated by modes with masses near $\TRH$\,. These modes decay to SM particles via Eq.~\eqref{eq:KK_graviton_coupling} with lifetimes of order $\tau \sim \mathcal{O}(10^{6{\text{--}}7}~{\rm yrs})$. To survive as DM, these heavy modes must instead \emph{fragment} to lower KK levels through intra-tower decays, leaving a present-day mass distribution near the keV scale. Such decays naturally outpace decays to the SM, owing to the large number of open decay channels. Intra-tower decays are more model dependent than the production process and result in a variety of interesting phenomenological probes~\cite{Mohapatra:2003ah,Dienes:2011ja,Dienes:2011sa,Dienes:2012jb,Law-Smith:2023czn} that strongly constrain the dark dimension scenario. 

A significant tension arises: if the masses become too light, dark matter becomes too warm to remain in DM halos, but if the masses remain too heavy decays to the SM can not comply with cosmological bounds on energy injection. Specifically, the simplest dark dimension dark matter model is excluded. Tuning the graviton-SM coupling of Eq.~\eqref{eq:KK_graviton_coupling}, the inter-tower decay rates, and the mass spectrum by hand can evade these bounds~\cite{Obied:2023clp,Law-Smith:2023czn}, but no concrete construction achieving this has been proposed.

In this letter, we extend the dark dimension scenario with $N$ axion KK towers motivated by the string axiverse~\cite{Svrcek:2006yi,Arvanitaki:2009fg,Cicoli:2012sz}. Such constructions generically have $N\gg 1$, with most axions effectively decoupled from SM gauge bosons~\cite{Gendler:2023kjt}. We assume throughout that only one axion couples to the SM. We show that independently of the initial distribution of energy among towers, they equilibrate at late times to a fixed distribution. When most axion towers couple to the SM only gravitationally, DM resides predominantly in these \textit{dark towers}, and energy injected into the SM is suppressed by $\mathcal{O}(1/N)$. This is a concrete mechanism by which dark dimension DM can evade phenomenological bounds. We do not address how the dark dimension is stabilized; moduli stabilization may impose further constraints, as discussed recently in \cite{Braun:2026amc}.

We first present the string-theoretic motivation for an axiverse within the dark dimension scenario in Sec.~\ref{sec:string theory}. The present-day abundance then separates cleanly into two parts. The \emph{freeze-in production} of each mode, dominated by inverse decay, is computed in Sec.~\ref{sec:Tower Freeze-In Production}. The subsequent \emph{fragmentation evolution}, driven by decays of tower states into one another, follows in Sec.~\ref{sec:Tower Fragmentation}. Its impact on dark dimension constraints appears in Sec.~\ref{sec:Phenomenological Bounds}. We conclude in Sec.~{\ref{sec:discussion}}.

\section{String Theory Motivation for Axion Towers} \label{sec:string theory}
It is natural to ask how the embedding of this scenario, an effective 5D theory with a large extra dimension, into a UV-complete description may lead to a richer structure (see e.g.,~\cite{Schwarz:2024tet,Heckman:2024trz,Reig:2025dpz,Blumenhagen:2026rgu} in the context of GUTs), especially with respect to the emergence of additional KK towers. Famously, within string theory, it is quite generic that axions arise at low energies as zero modes of different higher-dimensional gauge fields upon dimensional reduction \cite{Witten:1984dg,Choi:1985je,Svrcek:2006yi,Arvanitaki:2009fg,Cicoli:2012sz}. These include the Kalb-Ramond $B-$field as well as the Ramond-Ramond $p-$form gauge fields, $C_p$, which arise in theories with $D-$branes. The number of axions is typically given by the number of inequivalent cycles in the compact space (e.g., the Hodge number $h^{1,1}$ for $C_4$ axions in type IIB string theory) and can be $\sim O(100)$~\cite{Demirtas:2018akl} in type IIB constructions and up to $\sim O(10^{4-5})$ in F-theory models~\cite{Fallon:2025lvn}.

Consider a stack of $D(p+3)-$branes wrapping a $p-$cycle.
The Chern-Simons (CS) part of the $D-$brane action couples $C_p$ to gauge fields~\cite{Ibanez:2012zz}:
\begin{equation}
    S_{CS}\supset \mu_{p+3}\frac{(2\pi\alpha^\prime)^2}{2} \int C_p\wedge \text{tr}\,F^2 \,.
    \label{eq:Cp-coup}
\end{equation}
Here, $2\pi \sqrt{\alpha^\prime} = l_s$ and $\mu_{p+3}=\frac{(\alpha^\prime)^{-(p+4)/2}}{(2\pi)^{p+3}} $ are the string length and brane tension, and the integral is over the worldvolume of the $D$-branes (which we take orthogonal to the dark dimension). 
Assuming no other effect, such as fluxes, removes these fields from the low-energy spectrum, CS  interactions lead to axions coupled to gauge bosons in the 4D effective field theory, which obtain masses only from non-perturbative effects.
If one of these axions couples to QCD, it may provide a solution to the strong CP problem \cite{Peccei:1977hh,Peccei:1977ur,Weinberg:1977ma,Wilczek:1977pj}. Moreover, as shift-symmetry breaking effects are exponentially suppressed---a feature that is naturally explained in terms of global higher-form symmetries~\cite{Craig:2024dnl}---a stringy QCD axion naturally leads to a solution to the axion quality problem~\cite{Georgi:1981pu,Kamionkowski:1992mf,Barr:1992qq,Holman:1992us}.

The interaction strength of the axion is parametrized by the decay constant, $f_a$, which can be obtained from the kinetic term of the $C_p$ field,
\begin{align}
    S_{\rm kin}=-\frac{1}{2}\frac{2\pi}{l_s^{8-2p}}\int F_{p+1}\wedge\star F_{p+1}\,.
\end{align}
Its magnitude depends on the geometry of the compactified dimensions. Assuming an isotropic compact space, it is expected to be $f_a \sim M_P\left(\ell_s^p/\mathcal{V}_p\right)$, where $\mathcal{V}_p/\ell_s^p$ is the volume of the cycle in units of the string length~\cite{Svrcek:2006yi}. In more complicated geometries (e.g., non-factorizable geometries), the decay constant can be lowered and usually takes values near the string scale, $f_a \sim M_s$, which also decreases as the number of cycles increases. This has been verified explicitly for a class of type IIB string compactifications~\cite{Gendler:2023kjt,Benabou:2025kgx}; see also~\cite{Reece:2025thc,Reece:2026hmp} for recent studies.
The axion decay constant is expected to lie within
\begin{equation}\label{eq:decay_constant_range}
    M_s \lesssim f_a \lesssim \frac{\alpha_{\rm SM}}{2\pi}M_P\,,
\end{equation}

Integrating over the small compact dimensions, the interaction of the axion $a$ with the SM gauge bosons is 
\be
    \mathcal{L}\supset \sum_i \int dy \frac{ \kappa_i\, \alpha_i}{8\pi f_a}a(x,y) F_i\tilde{F_i}\delta(y)\,,
    \label{eq:5Dcoup}
\ee
where $\alpha_i$ is a SM gauge coupling,  $\kappa_i$ is an integer, and we left explicit the integral over the dark dimension, $y$.
Although axion modes can arise as bulk fields via the interaction in Eq.~\eqref{eq:Cp-coup}, other string theory constructions could localize them on the SM brane. As shown in~\cite{Gendler:2024gdo}, in this case, the decay constant is predicted to lie near the 5D Planck scale around $f_a\sim 10^{9-10}$ GeV, with important implications for QCD axion searches. Here, we focus on characterizing the phenomenology of axions from the closed string spectrum that propagate throughout the dark dimension and, therefore, have KK towers similar to the graviton. An example of such axion towers can arise in the context of heterotic M-theory with gauge sectors living in a 10D brane at the end of the interval, which could play the role of the dark dimension~\cite{Schwarz:2024tet,Blumenhagen:2026rgu}, provided proton decay can be kept under control~\cite{Reig:2025dpz}. Axions come in this case from the 3-form field $C$ and couple to gauge bosons via the CS interaction in the effective 11D action (see \cite{Svrcek:2006yi} for details). Once we integrate out the small dimensions of the CY at the end of the interval, we obtain an axion coupled to gauge bosons with a KK tower along the dark dimension.

\begin{figure*}[t]
    \centering
    \includegraphics[width=1\linewidth]{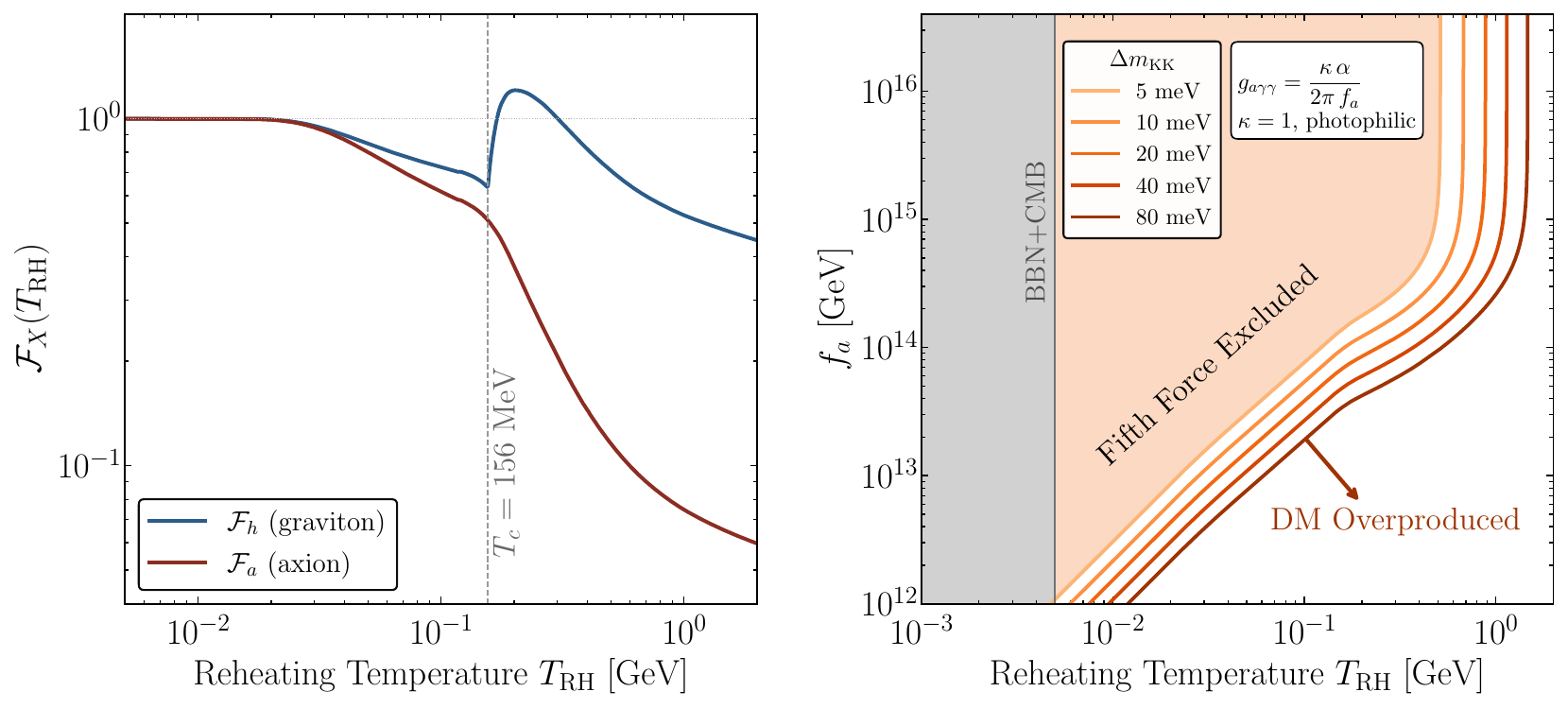}
    \caption{\textbf{Left:} freeze-in factors $\mathcal{F}_X(\TRH)$ defined by \Eq{eq:graviton_abundance} and \Eq{eq:axion_abundance}; see App.~\ref{app:Freeze In Abundance} for details. \textbf{Right:} contours showing values of $f_a$ and $\TRH$ required to obtain the observed DM abundance today for several $\mKK$. The oranged shaded region shows where fifth force bounds exclude $\mKK\lesssim 5$~meV.}
    \label{fig:Freeze-in}
\end{figure*}

Upon identifying the $y$-dimension in Eq.~\eqref{eq:5Dcoup} with the large~\textit{dark dimension} and integrating it out, an axion KK tower arises with the same mass scaling as the graviton tower. Similar to~\eqref{eq:KK_graviton_coupling}, KK modes of closed string axions interact with SM gauge bosons via
\begin{equation}
\mathcal{L}\supset{\sum_{i,n}\frac{\kappa_i\alpha_i}{8\pi f_a}a^{(n)}(x)F_i\tilde F_i }\,,
\end{equation}
with $n$ labeling KK number and $i$ parameterizing SM gauge sectors (there could be interactions with additional gauge sectors, which we ignore for simplicity).  This interaction suggests that the axion KK modes have couplings suppressed by $f_a$ but, similar to the graviton, receive enhancement to their freeze-in abundance from the large number of accessible modes. As their interaction with SM fields can be parametrically stronger than gravity (see Eq.~\eqref{eq:decay_constant_range}) the axion KK modes produced via freeze-in after reheating can contribute significantly to the DM abundance.

In addition to the axion coupled to the SM gauge bosons via the interaction in Eq.~\eqref{eq:5Dcoup}, dark dimension models can provide additional axion towers that do not couple sizeably to the SM. These can arise, for example, due to having many inequivalent cycles in a compact space with complicated topology (e.g. models with large $h^{1,1}$ in type IIB constructions). Some cycles may be ``far away'' from our visible brane, localized in the dark dimension, meaning that SM gauge bosons do not couple to these axions via CS interactions. Kinetic mixing of axions is also typically suppressed in this case, since separated cycles tend to not intersect; see, e.g.,~\cite{Gendler:2023kjt}. We refer to these as \emph{dark axion} towers; they play no role in the initial freeze-in production or decays into SM particles but will be important when understanding how the energy density is distributed through the different towers.


\section{Tower Freeze-In Production} \label{sec:Tower Freeze-In Production}
We calculate bulk mode abundances assuming initial conditions where the bulk is empty and the SM brane has temperature low enough to not thermalize bulk modes. Brane fields populate the bulk through non-renormalizable operators, i.e. UV freeze-in~\cite{Hall:2009bx}.

Using notation $X=h,a$, and taking the continuum limit valid for $\TRH\gg\Delta m_{\rm KK}$ (i.e. $\sum_i\to\int dm/\Delta m_{\rm KK}$), we define the yield density 
\begin{align}\label{eq:yield_density}
    Y_X(m,t)\equiv n_X(m,t)/s(t)\,,
\end{align}
with $n_X(m,t)\,\frac{dm}{\Delta m_{\rm KK}}$ the number density of $X$ modes with masses in $[m,m+dm]$. Freeze-in at reheating gives
\begin{align}\label{eq:freeze-in_integral}
    Y_{X,{\rm RH}}(m) = \int_{m/\TRH}^{\infty} \tilde{g}(x)\frac{R_{X}(m,x)\,dx}{x\,H(x)\,s(x)}\,,
\end{align}
where we used $x=m/T$, $H=(\pi^2 g_*(T)T^4/90M_P^2)^{1/2}$,  $s=2\pi^2 g_{*,s}(T)T^3/45$, and $\tilde{g}(T) = 1+\frac{1}{3}\frac{d \log g_{*,s}}{d\log T}$. The inverse-decay production rate with symmetry factor $S$ is
\begin{align}
    R_{X}(m,x) \approx \frac{ m^2 K_1(x)}{32\pi^3 x}\sum_{1,2}\frac{\left|\mathcal{M}_{1+2\rightarrow X}\right|^2}{S}.
\end{align}

Decays within and between KK towers change this picture, as first studied in~\cite{Dienes:2011ja, Dienes:2011sa, Dienes:2012jb, Obied:2023clp} and described in detail in \cref{sec:Tower Fragmentation}. We expect such decays to be nearly mass conserving: approximate mass conservation follows from approximate translation symmetry along the dark dimension, and is phenomenologically required at late times since releasing too much kinetic energy makes DM too warm to remain in DM halos today. 

Therefore decays to lighter KK states are an approximate \emph{fragmentation process}; the total DM mass is approximately conserved. The total energy density in the $X$ tower today is then obtained by red-shifting alone
\begin{align}
    \rho_{X}(t_0) \approx s_0 \int_0^\infty m\, Y_{X,{\rm RH}}(m) \frac{dm}{\mKK}\,;
\end{align}
$s_0 \approx 2.2\times 10^{-38}~\GeV^3$ is the entropy density today~\cite{Planck:2018vyg} and we use subscript ``$0$'' for quantities today throughout.

We consider now a simplified scenario where a single axion tower couples only with photons through
\begin{align}
    &\mathcal{L}_{a,\,{\rm int}} \equiv -\frac{\ga }{4}aF_{\mu \nu}\tilde{F}^{\mu \nu},\qquad \ga=\frac{\alpha_{\rm em} \kappa}{2\pi f_a}\,,
\end{align}
with $\kappa = 1$ used throughout. The graviton and axion freeze-in abundance normalized is
\begin{align}
    &\frac{\rho_{h,0}}{\rho_{{\rm DM},0}} \approx  1.3 \times 10^{-6}\left(\frac{\Delta m_{\rm KK}}{{5~\rm meV}}\right)^{-1}\times\notag \\
    &\qquad  \qquad \quad ~ \left(\frac{\TRH}{5~{\rm MeV}}\right)^3\,\mathcal{F}_h\left(\TRH\right) \,, \label{eq:graviton_abundance}\\
    &\frac{\rho_{{a},0}}{\rho_{{\rm DM},0}} \approx    1.2\left(\frac{f_a}{10^{12}~{\rm GeV}}\right)^{-2} \left(\frac{\Delta m_{\rm KK}}{{5~\rm meV}}\right)^{-1}\times\notag \\
    &\qquad  \qquad \quad ~ \left(\frac{\TRH}{5~{\rm MeV}}\right)^3\,\mathcal{F}_a\left(\TRH\right) \,. \label{eq:axion_abundance}
\end{align}
Benchmark values for $\mKK$ and $\TRH$ are chosen because fifth-force searches require $\mKK \gtrsim 5 $~meV ~\cite{Tan:2016vwu,Lee:2020zjt} and successful BBN predictions require $\TRH\gtrsim 5$~MeV~\cite{Hasegawa:2019jsa,deSalas:2015glj}. $\mathcal{F}_X(\TRH)$ describes corrections to the above expression from additional relativistic degrees of freedom beyond photons, electrons, and neutrinos. Fig.~\ref{fig:Freeze-in} shows $\mathcal{F}_X(\TRH)$ and the lower bound on $f_a$ as a function of $\mKK$ and $\TRH$; see App.~\ref{app:Freeze In Abundance} for further details.

We note that the freeze-in abundance need not be set by the maximal temperature after inflation, but by that of the last epoch of entropy release. An early matter dominated era dilutes the DM abundance produced before or during it (App.~\ref{app:EMD}). This decouples the low reheating $\TRH\lesssim1$~GeV from the inflationary scale. 

Finally, let us also comment on the contribution from misalignment for the axion tower. This contribution depends on the dynamics of inflation and is more model dependent than the freeze-in contribution considered above. If the universe is reheated after inflation to a maximal temperature $\TRH\sim\mathcal{O}(1~{\rm GeV})$, then the misalignment abundance is negligible in many cases. This can be seen, for example, by taking the instantaneous reheating approximation. In this case, $H_{\rm inf}\sim T_{\rm RH}^2/M_P \sim 10^{-9}$~eV. Any axion (be it a KK mode or the zero mode) with mass satisfying $H_{\rm inf}^4\lesssim m_a^2f_a^2$ and $H_{\rm inf}\lesssim m_a$ will quickly roll towards the minimum during inflation and have a negligible misalignment contribution after reheating. Cases where the inflationary scale does not satisfy these conditions require further study, see e.g.~\cite{Graham:2018jyp,Reig:2021ipa}. A full discussion of inflation in dark dimension scenarios, however, is beyond the scope of this paper, so we assume throughout that misalignment can be ignored.


\section{Tower Fragmentation} \label{sec:Tower Fragmentation}

Let us first focus on the evolution of a single KK tower, which we take to be gravitons. \emph{If} the KK graviton modes could only decay to SM states and had a mass spectrum centered above $O(1)$ MeV, the dark dimension would not be cosmologically viable. To see this, consider the decay rate to SM states for a graviton with mass $m$~\cite{Han:1998sg}:
\begin{align}
    \Gamma_{h\rightarrow \rm SM}(m) \approx \frac{m^3}{640 \pi M_P^2}\left(4n_1 + n_{1/2}\right) \,,
\end{align}
where $n_1$ ($n_{1/2}$) is the number of vector (fermion) degrees of freedom lighter than the graviton and phase space factors are ignored. Modes with $m \gtrsim 1$ MeV, for which the $\gamma\gamma$, $e^+e^-$ and $\nu\bar{\nu}$ channels are open, have lifetimes
\begin{align}\label{eq:tau_h_l}
    \tau_h(m) \lesssim t_{\rm U} \left(\frac{m}{100~{\rm MeV}}\right)^{-3},\quad {\rm for }~m\gtrsim 1~{\rm MeV}\,,
\end{align}
with $t_{\rm U}$ the age of the universe. Dark matter with such short lifetimes is excluded by searches for decaying dark matter~\cite{Boyarsky:2006ag,Boyarsky:2007ay,Ng:2019gch,Roach:2019ctw,Foster:2021ngm,Roach:2022lgo,Calore:2022pks}. Therefore KK gravitons must decay to lighter states within the tower to be DM. If such decays are allowed, their rates are much faster than decays to the SM due to the existence of many decay channels.

We neglect slow decays to the SM and model the total decay rate of a mode of mass $m$ as $\Gamma(m) = m^p/m_\star^{p-1}$, with $m_\star$ a reference scale set by KK-number-violating interactions. For near mass-conserving fragmentation through dimension-five operators, each of the $\sim m/\Delta m_{\rm KK}$ channels has a rate $\propto m^{5/2}$ (a factor of $m^{-1/2}$ arises from reduced phase space of near-degenerate final states) such that $p = 7/2$~\cite{Gonzalo:2022jac,Obied:2023clp}.

\subsection{Single Tower Evolution}

The yield density, defined in \Eq{eq:yield_density}, evolves as
\begin{align} \label{eq:y_m_t_equation}
    \dot{Y}(m,t) = &-\Gamma(m)\, Y(m,t) \notag 
    \\
    &+ \int_m^\infty \Gamma(m')F(m',m)\, Y(m',t)\,dm'.
\end{align}
The kernel $F(m',m)$ describes the average number of modes of mass $m$ produced in decays of modes of mass $m'$. Near-perfect mass conservation requires
\begin{align} \label{eq:mass_conservation_condition}
    \int_0^{m'} m\,F(m',m)\,dm = m'\,, \quad \left[{\rm mass~conservation}\right]
\end{align}
Requiring decays to two final-state particles gives
\begin{align} \label{eq:1_to_2_condition}
    F(m',m) = F(m',m'-m)\,,\quad \left[1 \to 2~{\rm decays}\right]
\end{align}
If $\Delta m_{\rm KK}\ll m'$, 
then $m'$ is the only kinematic scale in decays and fragmentation is homogeneous:
\begin{align}
    F(m',m) = \frac{1}{m'}\,g \left(\frac{m}{m'}\right) \quad \left[\rm one~scale\right].
    \label{eq:gcond}
\end{align}
Mass conservation and $1\to 2$ decays respectively require
\begin{align}
    \int_0^1 x\,g(x)\,dx = 1\quad{\rm and} \quad g(x) = g(1-x)\,.
    \label{eq:gproperties}
\end{align}
Using these assumptions and defining dimensionless variables $\mu = m/m_\star$ and $\tau = m_\star t$, the evolution equation is
\begin{align} \label{eq:y_m_t_equation_dimless}
    \dot{Y}(\mu,\tau) = &-\mu^p\, Y(\mu,\tau) \notag \\
    &+ \int_\mu^\infty \,\mu^{\prime\, p-1}\,g\left(\frac{\mu}{\mu'}\right) Y(\mu',\tau)\,d\mu'\,.
\end{align}
Invariance under $\mu\rightarrow \lambda \mu$ and $\tau\rightarrow \lambda^{-p} \tau$ singles out a scale-invariant attractor solution 
\begin{align}\label{eq:Y_att}
    Y(\mu,\tau) \underset{\tau\rightarrow \infty }{\longrightarrow} Y_{\rm att}(\mu,\tau)  = \mu^{-2}\,\Psi(\mu^p \tau)\,.
\end{align}
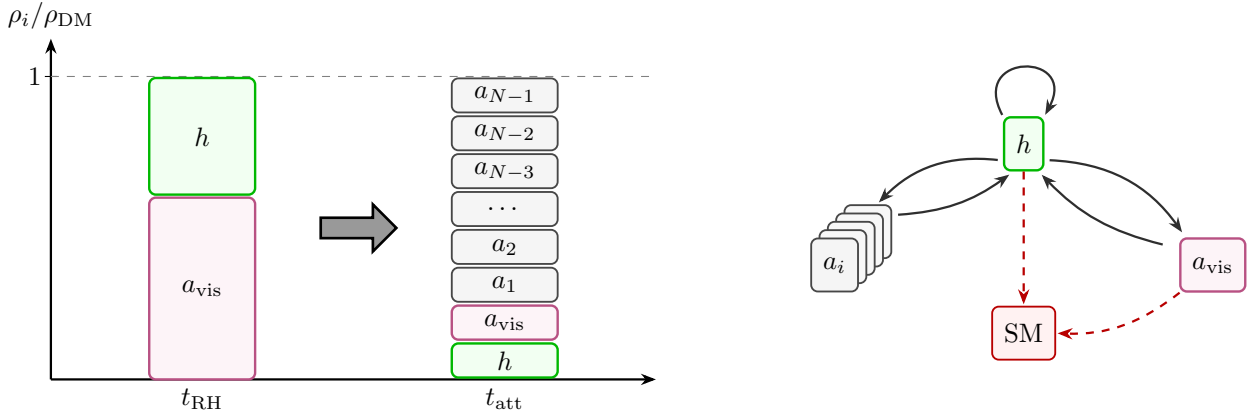
\begin{figure*}[t]
\centering
\begin{minipage}[c]{0.55\linewidth}
\centering
\begin{tikzpicture}[
    >={Stealth[length=2mm]},
    res/.style={
        draw, rounded corners=2.5pt, fill=black!4, draw=black!72,
        line width=0.7pt, text width=1.4cm, minimum height=0.45cm,
        align=center, inner sep=0pt
    },
    dark/.style={->, line width=0.85pt},
    grav/.style={res, fill=green!5, draw=green!72!black, line width=0.9pt},
    qcd/.style={res, fill=magenta!5, draw=magenta!72!black, line width=0.9pt}
]
\draw[dark] (0,0) -- (8,0) node[right] {};
\draw[dark] (0,0) -- (0,4.5) node[above] {$\rho_i/\rho_{\rm DM}$};

\node[left] at (0,4) {$1$};
\draw (-0.08,4)--(0.08,4);
\draw[dashed, gray] (0,4)--(8,4);

\node[below] at (2,0) {$t_{\rm RH}$};
\node[below] at (6,0) {$t_{\rm att}$};
\node[single arrow,
    fill=black!40, draw=black!100, line width=1.0pt,
    minimum width=12pt, single arrow head extend=5pt,
    minimum height=10mm] at (4,2) {};

\node[draw, rounded corners=2.5pt, fill=magenta!5, draw=magenta!72!black,
      line width=0.9pt, minimum width=1.4cm, minimum height=2.4cm] at (2,1.2) {$a_{\rm vis}$};
\node[draw, rounded corners=2.5pt, fill=green!5, draw=green!72!black,
      line width=0.9pt, minimum width=1.4cm, minimum height=1.54cm] at (2,3.21) {$h$};

\node[grav] at (6,0.25) {$h$};
\node[qcd]  at (6,0.75) {$a_{\rm vis}$};
\node[res]  at (6,1.25) {$a_1$};
\node[res]  at (6,1.75) {$a_2$};
\node[res]  at (6,2.25) {$\dots$};
\node[res]  at (6,2.75) {$a_{N-3}$};
\node[res]  at (6,3.25) {$a_{N-2}$};
\node[res]  at (6,3.75) {$a_{N-1}$};
\end{tikzpicture}
\end{minipage}%
\hfill
\begin{minipage}[c]{0.43\linewidth}
\centering
\begin{tikzpicture}[
    >={Stealth[length=2mm]},
    res/.style={draw, rounded corners=2.5pt, inner sep=4.5pt, minimum height=7mm,
                align=center, line width=0.7pt, fill=black!4, draw=black!72},
    grav/.style={res, fill=green!5, draw=green!72!black, line width=0.9pt},
    qcd/.style={res, fill=magenta!5, draw=magenta!72!black, line width=0.9pt},
    smbox/.style={res, fill=red!5, draw=red!72!black},
    dark/.style={->, line width=0.85pt, black!82, shorten >=0.5pt, shorten <=0.5pt},
    sm/.style={->, line width=0.85pt, dashed, red!72!black, shorten >=0.5pt}
    ]
    \coordinate (H) at (0,1.5);    
    \coordinate (A) at (-2.5,-0.1);
    \coordinate (Q) at (2.5,-0.1); 
    \coordinate (S) at (0,-1.0);   

    \node[grav] (h) at (H) {$h$};

    \begin{scope}[on background layer]
    \foreach \i in {4,3,2,1} {
      \node[res, fill=black!3] at ($(A)+(\i*1.1mm,\i*1.1mm)$) {\phantom{$a_i$}};
    }
    \end{scope}
    \node[res] (a) at (A) {$a_i$};

    \node[qcd] (q) at (Q) {$a_{\mathrm{vis}}$};
    \node[smbox] (s) at (S) {SM};

    \draw[dark] (h.north west) .. controls +(-0.45,0.85) and +(0.45,0.85) .. (h.north east);
    \draw[dark] ($(h.south west)+(-0.5mm,1.5mm)$) to[bend right=26] ($(a.north east)+(2.5mm,4.5mm)$);
    \draw[dark] ($(a.north east)+(5mm,3mm)$)     to[bend right=16] ($(h.south west)+(1.mm,-.3mm)$);
    \draw[dark] ($(h.south east)+(0.5mm,1.5mm)$)  to[bend left=26] ($(q.north west)+(0.5mm,0.5mm)$);
    \draw[dark] ($(q.north west)+(-2mm,-1mm)$)    to[bend left=16] ($(h.south east)+(-0.5mm,-.3mm)$);
    \draw[sm] (h) -- (s);
    \draw[sm] (q) to[bend left=20] (s);
    \end{tikzpicture}
\end{minipage}
\caption{\textbf{\emph{Left:}} Schematic description of multi-tower evolution. At late times (past the attractor time $t_{\rm att}$) most of the dark matter, initially produced as graviton (green) and a tower of visible KK axions coupled (magenta), resides in the dark axion towers (in gray). \textbf{\emph{Right:}} Fragmentation connecting the towers and the Standard Model (dashed red).}
\label{fig:Fragmentation_diagram}
\end{figure*}
The weight $\mu^{-2}$ is fixed by mass conservation and $\Psi(\xi)=d\mathcal{M}/d\ln\mu$ is mass per e-fold in $\mu$, where
\begin{align}
    \mathcal{M}=\int_0^\infty \mu\, Y(\mu,\tau)\,d\mu
\end{align}
is the conserved total mass of the tower (in units of $m_\star$). This scaling also fixes the average mass of the distribution to evolve proportional to $\tau^{-1/p}$. Using the scaling invariant $\xi\equiv \mu^p \tau$, the attractor satisfies
\begin{align}
    &\Psi'(\xi)  = - \Psi(\xi) +  G(\xi)\,,\\
    &G(\xi) \equiv \frac{\xi^{\frac{2}{p}-1}}{p}\int_\xi^\infty g[(\xi/\xi')^{1/p}]\,\xi^{\prime -\frac{2}{p}}\,\Psi\left(\xi'\right)d\xi'\,, \label{eq:G}
\end{align}
where the $-\Psi(\xi)$ describes decays at $\xi$ and $G(\xi)$ describes decays of modes with $\xi' >\xi$ to $\xi$. It has the solution
\begin{align} \label{eq:f_sol}
    \Psi(\xi) = e^{-\xi}\int_0^\xi e^{\xi'}G(\xi')\,d\xi'\,.
\end{align}
This can be solved numerically by the following method:
\begin{enumerate}
    \item Choose a trial solution $\Psi_1(\xi)$.
    \item For $i\geq 1$ evaluate $G_i(\xi)$ using $\Psi_i(\xi)$ and solve for $\Psi_{i+1}(\xi)$ using \cref{eq:f_sol}; normalize to conserve mass.
    \item Repeat step 2 until a threshold is reached.
\end{enumerate}
Results of this method are shown in Fig.~\ref{fig:iterations}. To test this method, we compare with the exactly solvable solution
\begin{align}
    \Psi_{\rm uniform}(\xi) = \dfrac{p\mathcal{M}}{\Gamma\left(\frac{2}{p}\right)}\,\xi^{\frac{2}{p}}\,e^{-\xi}\,, \quad {\rm for}~g(x)=2,
    \label{eq:fdist}
\end{align}
where we define the conserved total mass of all modes
\begin{align} \label{eq:mass_conservation}
    \mathcal{M} = \int_0^\infty \mu\, Y(\mu,\tau) \, d\mu\, = \frac{1}{p}\int_0^\infty \Psi(\xi)\,d\log \xi\,.
\end{align}


Initial conditions, and therefore solutions at generic times, need not satisfy this scaling symmetry. However, the distribution approaches the attractor solution at late times regardless of  initial conditions. Here ``late'' means that heavy modes with masses near $\TRH$, which initially dominate the energy density, have decayed. An analytic demonstration of this is given in \ref{app:Calculation of Tower Fragmentation}. 
The time to reach the attractor can be expressed through the peak of the mass distribution today, $m_0$, as
\begin{align}
    t_{\rm att} \sim t_{\rm U} \left( \frac{m_{0}}{\TRH}\right)^p\,.
    \label{eq:tasymp}
\end{align}
At times later than $t_{\rm att}$ the distribution is well approximated by \Eq{eq:Y_att}. In particular, the attractor form already holds at recombination provided $m_0\lesssim 0.05\, \TRH$, which is always the case for our purposes given \Eq{eq:tau_h_l}.


\subsection{Multi-Tower Evolution}

We now consider one graviton tower $h$ and $N$ axion towers $a_1,\dots,a_N$. We solve for the attractor solution of the fragmentation process (sketched in \Fig{fig:Fragmentation_diagram}) and show energy injection to the SM is reduced by a factor of $N$.
\begin{figure*}[t]
    \centering
    \includegraphics[width=1\linewidth]{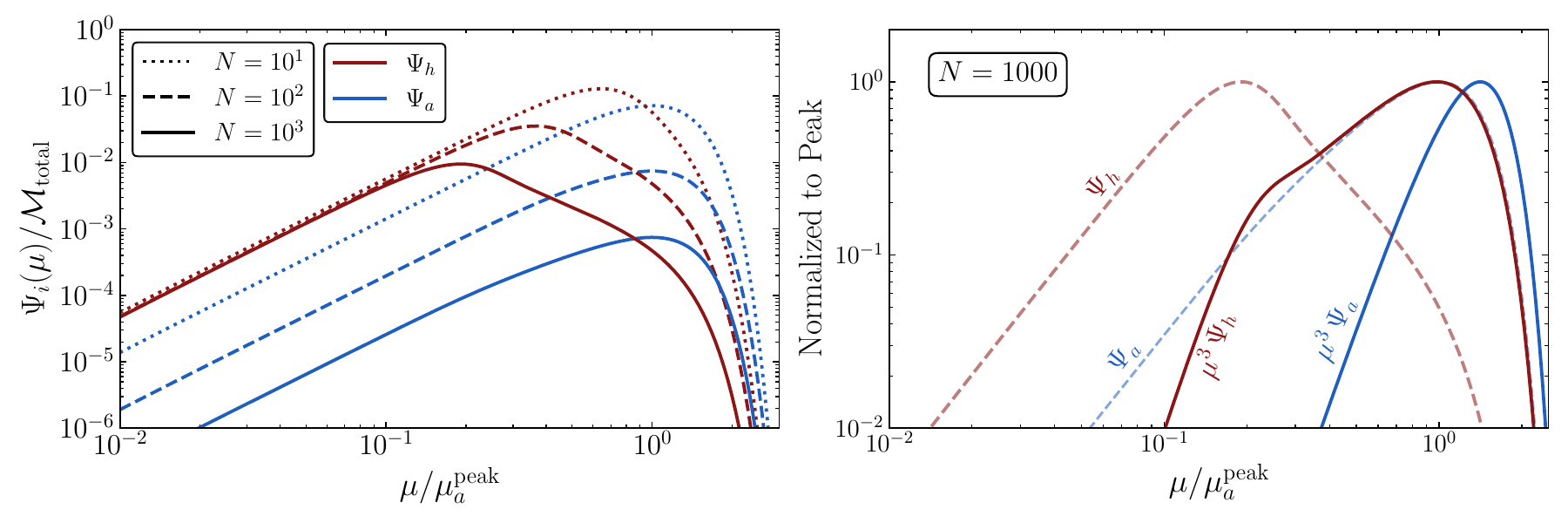}
    \caption{\textbf{\emph{Left:}} Numerical solutions for the attractor mass distributions of graviton towers (red) and single axion towers (blue) for $N=10,\,10^2,\,10^3$ ($\alpha=1$, $g(x) = 2$), normalized to total mass. Increasing $N$ suppresses both towers and moves the graviton peak to smaller masses $\propto N^{-1/p}$. Above its peak the graviton is still sourced by axions and cuts off only above the axion peak. \textbf{\emph{Right:}} Mass distributions at $N=10^3$ (dashed) compared to energy injection rates $\propto \mu^3\Psi_i$ (solid), each normalized by its own peak value. This demonstrates energy injection arises primarily from masses near the axion mass peak for both towers.}
    \label{fig:distributions}
\end{figure*}

Yield densities, $\bm{Y} = (Y_h,\,Y_1,\dots,Y_N)^\top$, evolve as 
\begin{align}
    \dot{\bm{Y}}(\mu, \tau) = &-\mu^p\, \bm{C}\cdot\bm{Y}(\mu, \tau) \\
    \label{eq:y_m_t_equation_2D}
    &+ \bm{F}\cdot\bm{C}\cdot\int_\mu^\infty  \mu^{\prime\, p-1} g\left(\frac{\mu}{\mu'}\right) \,\bm{Y}(\mu',\tau)\,d\mu'\,,\nonumber
\end{align}
where for simplicity we use a single fragmentation kernel $g$ for all channels (in general the processes $h\to hh$, $h\to a_i a_i$, and $a_i\to a_i h$ need not follow this assumption, which we make for simplicity). 
We assume gravitons couples equally to all axions and that axion towers do not decay into one another.
The total graviton decay rate scales as $\sim \mathcal{O}(N^1)$ while the total axion decay rate scales with the number of towers as $\sim \mathcal{O}(N^0)$. We approximate the graviton and axion decay rates as 
\begin{align}
    &\Gamma_i(m) =\Lambda_i\,m^p/m_\star^{p-1}\,,
\end{align}
where $\Lambda_h\sim 1+N\alpha$ and $\Lambda_a \sim \alpha$, and $\alpha$ is an $O(1)$ ratio of the $a^2 h$ coupling to the $h^3$ coupling. This gives
\begin{align}
    \bm{C} = {\rm diag}(\Lambda_h ,\Lambda_a,\dots,\Lambda_a)\,.
\end{align}
The generalized mass conservation condition,
\begin{equation}
    \frac{\partial}{\partial \tau} \left(\sum_i\int \mu \mathbf Y_i(\mu,\tau) \, \text{d}\mu\right) = 0\,,
\end{equation}
implies
$\sum_i F_{ij} = 1$ for all $j$ such that
\begin{align} \label{eq:F0_multi}
    \bm{F} = \begin{pmatrix}
        \dfrac{1}{\Lambda_h} & (1/2)\,\bm{1}_N^{\top} \\[12pt]
        \dfrac{\Lambda_a}{\Lambda_h} \,\bm{1}_N & (1/2)\,\bm{I}_N
    \end{pmatrix},
\end{align}
where $\bm{I}_N$ is the $N\times N$ identity and $\bm{1}_N^\top = (1,\,...,1)$. 
We again solve for the attractor solution as scale invariance still exists. We consider solutions of the form
\begin{align}
    &\bm{Y}(\mu,\tau) = \mu^{-2}\,\bm{\Psi}(\mu^p\tau)\,.
\end{align}
The equations determining attractor solutions are
\begin{align}
    & \bm{\Psi}'(\xi)  = -\bm{C}\cdot \bm{\Psi}(\xi) + \bm{G}(\xi) \,, \label{eq:diffeq_u_multi}\\
    &\bm{G}(\xi) \equiv 
    \frac{\xi^{\frac{2}{p}-1}}{p}\bm{F}\cdot\bm{C}\int_\xi^\infty g\left[\left(\xi/\xi'\right)^{\frac{1}{p}}\right]\xi^{\prime -\frac{2}{p}}\bm{\Psi}(\xi')d\xi'\, \label{eq:source_multi}.
\end{align}
The solution is 
\begin{align}
    \bm{\Psi}(\xi) &= e^{-\xi \bm{C}}\int_0^\xi e^{\xi'\bm{C}}\bm{G}(\xi')\,d\xi'\,, \label{eq:multi-tower_evolution}
\end{align}
and can be solved numerically by the iterative method described for the single tower; results are shown in Fig.~\ref{fig:distributions}.

The graviton distribution for $N\gg1$ is determined by
\begin{equation}
    \Psi_h^\prime \sim N\, \left(-\Psi_h + \frac{1}{2}\mathcal{G}[\Psi_a]\right)\,.
    \label{eq:simple}
\end{equation}
The source function in \Eq{eq:source_multi} scales as $\mathcal G[\Psi_a]\propto \xi^{\frac{2+s}{p}-1}$ for $\xi\ll1$ assuming $g(x)\sim x^s$ for $x\ll1$ and $s>-2$.

For $\xi\ll N^{-1}$, graviton modes haven't had time to decay such that the first term on the RHS of~\Eq{eq:simple} can be neglected, giving $\Psi_h' \approx\frac{N}{2} \mathcal G[\Psi_a]\implies \Psi_h\propto \xi^{\frac{2+s}{p}}$. 
%
For $ N^{-1}\ll \xi\ll1$, the LHS of~\Eq{eq:simple} can be neglected, giving $\Psi_h\approx\frac{1}{2} \mathcal G[\Psi_a] \implies \Psi_h\propto \xi^{\frac{2+s}{p}-1}$. 
%
Finally, modes with $\xi\gg 1$ are exponentially suppressed. 

The shape of axion tower mass distributions is similar to a single graviton towers with a shape modification from the fact that KK axion decays only produce one daughter KK axion. These behaviors are shown in Fig.~\ref{fig:distributions}.

We denote the mass in the graviton (a single axion) tower by ${\cal M}_h$ (${\cal M}_a$) using the analogue of \Eq{eq:mass_conservation} and define ${\cal M}_{\rm tot} = \mathcal{M}_h + N \mathcal{M}_a$. The above scaling fixes
\begin{align}
    &\mathcal{M}_a/\mathcal{M}_{\rm tot} \propto N^{-1},\\ &\mathcal{M}_h/\mathcal{M}_{\rm tot}  \propto N^{-{\rm min}(1,\,\frac{s+2}{p})}\quad {\rm for}~s>-2;
\end{align}
the minimum accounts for the transition between regimes where the mass is dominated near $\xi \sim \Lambda_h^{-1}$ vs $\xi \sim \Lambda_a^{-1}$.

Of more relevance to phenomenology is the moment of the mass distribution weighted by $\mu^3$ to account for the rate of decays to the SM which we define as
\begin{align}
    \mathcal{M}_3[\Psi]\equiv \int_0^\infty d\mu\, \mu^{3}\Psi(\mu).
\end{align}
For gravitons and axions, this is controlled by $\xi \sim \Lambda_a^{-1}$ and scale as $N^{-1}$ relative to the third moment of all DM.

\begin{figure*}[t]
    \centering
    \includegraphics[width=1\linewidth]{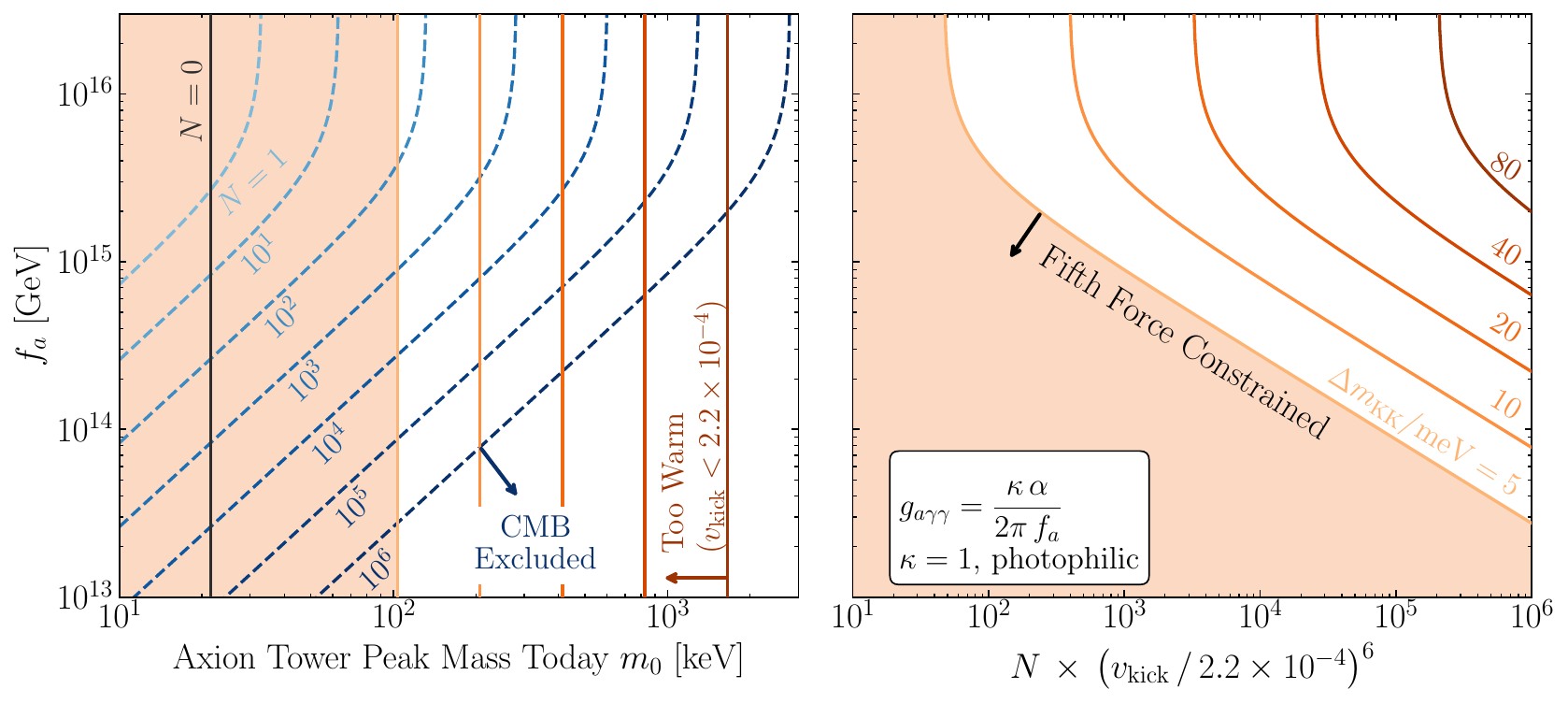}
    \caption{Late time constraints on dark dimension dark matter with $N$ axion towers assuming KK modes make up all of DM. \textbf{\emph{Left:}} Blue dashed contours show lower bounds on $f_a$ vs the axion-tower peak mass today $m_{0}$ for several values of $N$ from \Eq{eq:CMB_Bounds}. The $N=0$ curve is the CMB upper bound on the peak mass of an isolated graviton tower. Each additional tower dilutes the visibly decaying population and relaxes the bound. Solid orange contours show the upper bound on $m_0$ from warm DM bounds \Eq{eq:Warm_DM_Bounds}, for various $\Delta m_{\rm KK}$ (labeled in the plot on the right). The shaded region is excluded by the combination of the fifth force bound and the warm DM bound.
    \textbf{\emph{Right:}} Constraints on the left in terms of axion tower parameters, $f_a$ and $N$. }
\label{fig:fa_vs_mpeak}
\end{figure*}

\begin{figure*}
    \centering
    \includegraphics[width=1\linewidth]{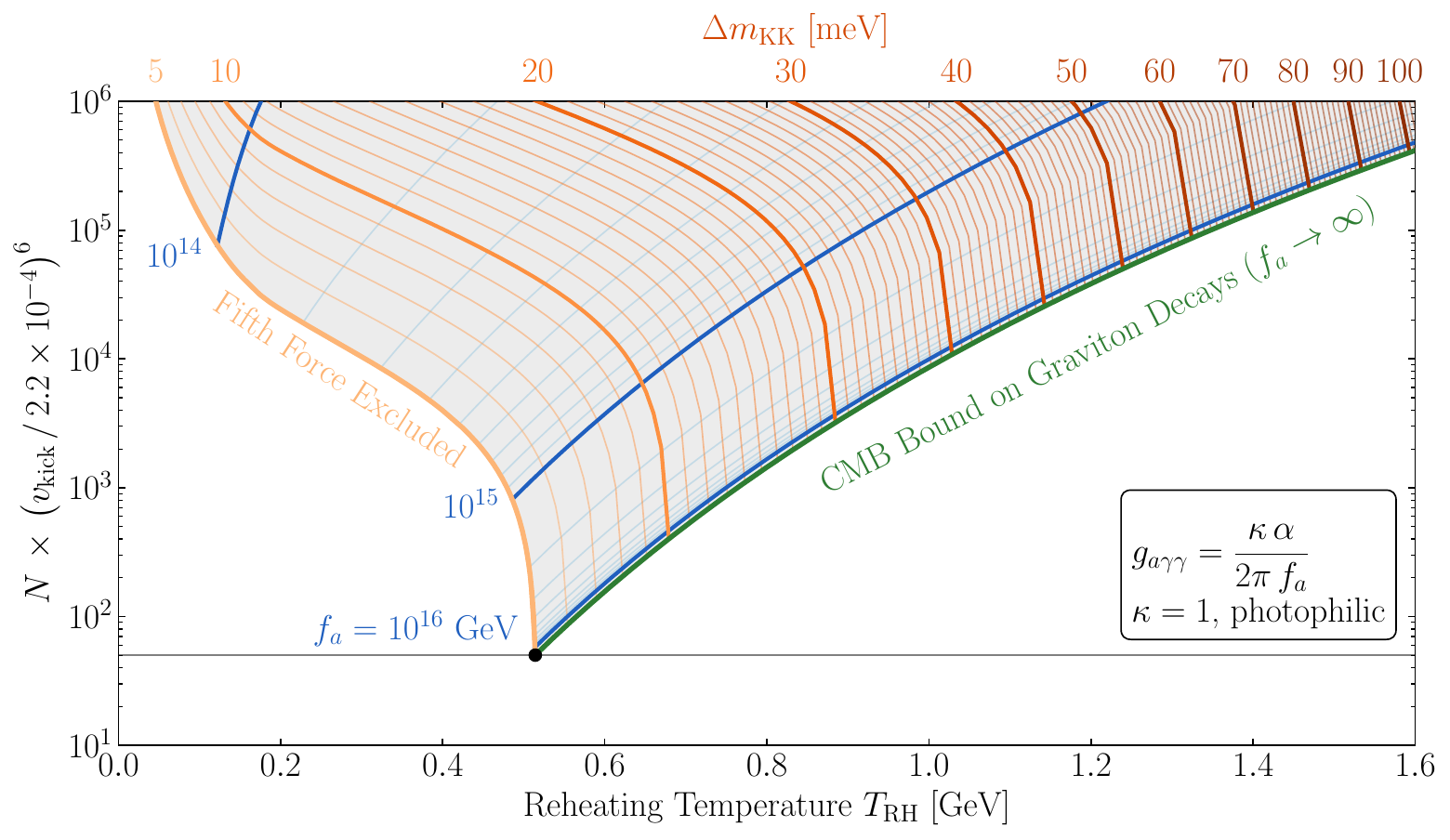}
    \caption{
    Allowed region (shaded) for dark dimension dark matter in the $(\TRH,N)$ plane. A point is viable if some $(\mKK,\,f_a)$ allows the towers to be all of dark matter ($\rho_a+\rho_h=\rho_{\rm DM}$, using  \Eq{eq:graviton_abundance} and \Eq{eq:axion_abundance}) while obeying the fifth-force floor $\mKK\gtrsim5$~meV, warm dark matter bounds from \Eq{eq:Warm_DM_Bounds}, and CMB energy injection bounds from \Eq{eq:CMB_Bounds}. Orange (blue) contours mark the largest (smallest) value of $\mKK$ ($f_a)$ consistent with these bounds as a function of $\TRH$ and $N$. The green contour shows the limit where the CMB bounds depend only on graviton decays, i.e. the limit where all axion towers are dark.
    }
    \label{fig:icecream_cone}
\end{figure*}

\section{Phenomenological Bounds}  \label{sec:Phenomenological Bounds}

This section describes how bounds on a single graviton tower are recast to give bounds on the multi-tower dark dimension scenario. We use approximate bounds given in~\cite{Obied:2023clp, Gonzalo:2022jac, Law-Smith:2023czn}; more precise bounds are beyond the scope of this paper as our aim is limited to showing how these bounds are modified by the presence of $N$ axion towers.

\emphb{Fifth Forces: } Tests of Newton's law of gravity constrain $\mKK\gtrsim 5$~meV~\cite{Tan:2016vwu,Lee:2020zjt,Obied:2023clp}. This bound only depends on the geometry of the extra dimension and is therefore independent of the particle content of the bulk.

\emphb{Warm DM: } 
Let $m_0$ denote the dark-matter mass distribution peak today; for $N\geq1$ this is the common peak of the axion towers, the graviton distribution peaking a factor $N^{-1/p}$ lower. Late time fragmentation releases kinetic energy from non-perfect mass conservation; assuming a KK tower with masses uniformly spaced in units of $\mKK$ requires decays to imparts a minimal ``kick'' velocity $v_{\rm kick}\approx\sqrt{\mKK/m_0}$ to daughter particles. Consistency with structure formation requires $v_{\rm kick}\lesssim 2.2\times10^{-4}$~\cite{Obied:2023clp} which requires
\begin{align} \label{eq:Warm_DM_Bounds}
    m_0 \gtrsim 100\,\keV \left(\frac{\mKK}{5\,\meV}\right)\left(\frac{v_{\rm kick}}{2.2\times 10^{-4}}\right)^{-2}.
\end{align}

\emph{\textbf {CMB Energy Injection:}} Dark matter decaying around recombination distorts CMB anisotropies. Analyzing Planck data, Ref.~\cite{Law-Smith:2023czn} evolves a graviton tower through this epoch and derives the following bound on an effective KK graviton coupling strength $\lambda$ multiplying \Eq{eq:KK_graviton_coupling} in terms of its peak mass today:
\begin{align} \label{eq:CMB_Bounds}
    \lambda^2\left(\frac{m_0}{100\,\keV}\right)^3 \lesssim 0.01.
\end{align}
To recast \Eq{eq:CMB_Bounds} for the multi-tower scenario we normalize energy injection to an isolated graviton tower $\Psi_{\rm iso}$ carrying all of the dark matter, with the same peak mass $m_0$ today. We define the effective injection weights
\begin{align} \label{eq:eps}
\varepsilon_{X}(N)\,\equiv\,
\mathcal{M}_3[\Psi_X]/\mathcal{M}_3[\Psi_{\rm iso}]
\end{align}
The multi-tower scenario injects  visible energy exactly as the benchmark graviton tower with effective coupling
\begin{align} \label{eq:lambda_eff}
    \lambda^{2}(N,f_a)
    = \varepsilon_h(N)\;+\;\frac{5}{4}\left(\ga M_P\right)^{2}\varepsilon_a(N)\,,
\end{align}
 where $\frac{5}{4}(\ga M_P)^2=\Gamma(a\to\gamma\gamma)/\Gamma(h\to\gamma\gamma)$.
The weights are where the fragmentation model enters. For the graviton tower and a single axion tower and both scale as $1/N$; for the axion this is because one axion tower shares only $1/N$ of the total axion tower energy density. For the graviton tower, most energy density sits at much smaller masses than $m_0$, but the third moment relevant for energy injection is sensitive only to the heaviest masses fed by the axion distribution (see Fig.~\ref{fig:distributions}); therefore also scaling as $1/N$. 

Specifically, for $N\gg1$ and $g(x)=2$ we have $\varepsilon_a(N)\approx 0.69/N$ and $\varepsilon_h(N)\approx 0.46/N$. Using these values, the minimal value of $N$ where both CMB bounds (\Eq{eq:CMB_Bounds}) and warm DM bounds (\Eq{eq:Warm_DM_Bounds}) are satisfied is 
\begin{align}
    N_{\rm min} = 50&\left(\frac{\mKK}{5\,{\rm meV}}\right)^3\left(\frac{v_{\rm kick}}{2.2\times 10^{-4}}\right)^{-6}\times \notag \\
    &\left[1+\left(\frac{f_a}{4\times 10^{15}\,{\rm GeV}}\right)^{-2}\right]\,.
\end{align}
\Fig{fig:fa_vs_mpeak} recasts these bounds in terms of the axion-tower parameters. This shows how improvements in fifth-force experiments that raise the lower bound on $\mKK$, or improvements that lower the upper bound on $v_{\rm kick}$, might further constrain $N$. As an example, improving bounds on $\mKK$ by a factor of $6$ or on $v_{\rm kick}$ by a factor of $2.4$ would already exclude most models with $N < 10^4$. This result would already put the multi-tower axiverse extension of dark dimension DM into serious tension with the ``reasonable'' string theoretic expections for the number of axion KK towers. \Fig{fig:icecream_cone} further synthesizes these results with freeze-in production results of Fig.~\ref{fig:Freeze-in} to show the viable parameter space in the $(\TRH,N)$ plane.

\emph{\textbf{Astrophysical bounds:}}  Stellar cooling bounds (see \cite{Hardy:2025ajb} for a recent study of KK gravitons), being independent of the DM fraction, could provide constraints on visible axion towers. Current bounds, assuming long-lived particles that do not get trapped within the star, constrain a single axion coupling to photons to be $(1/f_a)_{\rm single} \lesssim 10^{-5}\,\GeV^{-1}$~\cite{Lucente:2020whw}. In this scenario, for typical supernovae core temperatures, $T_\star\sim \mathcal{O}(10)$\,MeV, one expects around $\mathcal N\sim T_\star/\Delta m_{\rm KK} \sim 10^{10}$ KK axions to be efficiently produced. Re-interpreting the single axion bound on $\sqrt{\mathcal N}/f_a$ implies $f_a \gtrsim 10^{10}\,$GeV. This bound is weaker than the overproduction constraint.

\emph{\textbf{Axion towers, Casimir energy, and KK mass splitting:}}
An interesting implication of having $N$ axion towers whose zero mode mass is smaller than the KK mass splitting is their impact on the Casimir energy contribution to the vacuum energy~\footnote{We thank Miguel Montero for raising this point.}. In this limit, the 4D vacuum energy scales as
\begin{equation}
    V_{\rm{C}}(R)\sim \lambda_{\rm C}NR^{-4}= \lambda_{\rm C} N \Delta m_{\rm KK}^4\,,
\end{equation}
with $\lambda_{\rm C}$ a calculable coefficient. This implies that the limit $N\gg 1$ is equivalent to a moderate increase to the KK mass splitting, $\Delta m_{\rm KK}$. Equivalently, for a given compactification where the coefficient $\lambda_{\rm C}$ can be calculated (for example, $\lambda_{\rm C}=5\times 10^{-5}$ for a circle compactification, see~\cite{Montero:2022prj}) one could obtain upper bounds to $N$ by assuming that the Casimir contribution saturates the vacuum energy (assuming no cancellation from e.g. bulk fermions). This is an interesting direction that we leave for future studies.


\section{Conclusions}
\label{sec:discussion}

In this work we studied the impact of an axiverse on the phenomenology and cosmology of the dark dimension dark matter, considering
 closed string axions coupled to the Standard Model gauge bosons in the 4D theory. These axions can propagate in the bulk, forming KK towers analogous to that of gravitons.
 
%

The potentially stronger coupling of KK axions to the Standard Model implies their freeze-in production may naturally dominate the initial dark matter abundance, which can be explained for 
 $\mathcal{O}(1)\text{ MeV }\lesssim \TRH \lesssim \mathcal{O}(1)$ GeV and $10^{12} \text{ GeV }\lesssim f_a\lesssim 10^{16} \text{ GeV}$, as shown in Fig.~\ref{fig:Freeze-in}.
Even if the DM is dominated by the axion tower initially, 
decays of KK axions can produce KK gravitons at late times. In turn, these gravitons can decay and be sourced by additional \emph{dark} towers, decoupled from the Standard Model but necessarily coupled gravitationally.

We have studied numerically the impact of these processes on the species evolution, finding an attractor solution, established by the time of recombination, which simplifies phenomenological studies.  
%
%
%
We found that once the attractor solution is reached, all towers equilibrate, 
suppressing energy injection from visible towers by a factor of $\mathcal{O}(1/N)$ -- that is, by the number of KK towers. 
Critically, this implies phenomenological bounds weaken significantly for large $N$, opening up viable parameter space for dark dimension DM. 

Combining these with fifth-force bounds, we found that a minimal number of towers $N\gtrsim 50$ is required (Fig.~\ref{fig:icecream_cone}) for dark dimension dark matter to be viable without the need to modify couplings or spectra by hand. 
An $O(1)$ sensitivity improvement in these experiments---pushing the lower bound on $\Delta m_{\rm KK}$ to larger values---could strengthen the bound to $N\gtrsim 10^4$.
From a top-down perspective, this would have important implications. For example, compactifications based on Calabi-Yau threefolds that belong to the so-called Kreuzer-Skarke list have up to $N=491$ axions~\cite{Demirtas:2018akl}.
In the case of F-theory compactifications, there are ensembles with $N\gg 50$. However, in the geometric regime (where cycles are larger than 1 in string units), such a large number of cycles implies that the string scale decreases well below the Planck scale. In the ensembles studied in \cite{Fallon:2025lvn}, the typical decay constant for $N\sim 10^4$ would lie around $f_a\sim 10^{8-9}$ GeV or even smaller. If such axion propagates through the dark dimension the associated freeze-in abundance (see Eq.~\eqref{eq:axion_abundance}) would be overproduced by a factor $\sim 10^8$ even for the lowest reheating temperatures compatible with BBN, $T_{\rm RH}\sim 5$~MeV. These theoretical considerations set a direct testable constraint on the dark dimension scenario studied here.

More broadly, the analytic equilibration framework derived here applies beyond this specific setup, and can inform other phenomenological studies of KK tower dynamics.

\vspace{0.5 cm}
\noindent {\it Acknowledgments.} We thank Andrea Caputo, Miguel Montero, Georges Obied and
Irene Valenzuela for insightful discussions. M.R. and M.R. acknowledge support from the COST Action ”Cosmic
WISPers in the Dark Universe: Theory, astrophysics, and experiments” (CA21106).


\bibliographystyle{apsrev4-1}
\bibliography{bibliography}


\clearpage
\onecolumngrid
\begin{center}
   \textbf{\large SUPPLEMENTARY MATERIAL \\[.2cm] ``The Dark Dimension meets the Axiverse''}\\[.2cm]
  \vspace{0.05in}
  {Kevin Langhoff, Maria Ramos, and Mario Reig}
\end{center}

\setcounter{equation}{0}
\setcounter{figure}{0}
\setcounter{table}{0}
\setcounter{section}{0}
\setcounter{page}{1}
\makeatletter
\renewcommand{\theequation}{S\arabic{equation}}
\renewcommand{\thefigure}{S\arabic{figure}}
\renewcommand{\thetable}{S\arabic{table}}
\renewcommand{\thesection}{S\Roman{section}}

\section{Calculation of Freeze-In Relic Abundance}
\label{app:Freeze In Abundance}

This appendix calculates the freeze-in abundance quoted in \eqref{eq:graviton_abundance} and \eqref{eq:axion_abundance}. A mode of mass $m$ evolves according to the Boltzmann equation $\dot n + 3Hn = R(T)$. The bulk stays far below equilibrium, so the collision term keeps only production. Writing $Y_X=n/s$ and $x=m/T$ turns this into the yield integral \eqref{eq:freeze-in_integral}, with $\tilde g$, $H$, and $s$ as defined there.

The dominant process by far is inverse decay of the form $1+2\to X$; all kinematic invariants are fixed such that $|\mathcal{M}|^2$ is constant and can be pulled out of the phase-space integral. In the freeze-in limit, the statistical weight reduces to $f_X^{\rm eq}\left[1\pm(f_1^{\rm eq}+f_2^{\rm eq})\right]$ ($+$ bosons, $-$ fermions)~\cite{Langhoff:2022bij}. With $m_1=m_2$ the rate integrates exactly to
\begin{align}
\label{eq:R_ID}
R_{\rm ID}(m, T)=\sum_{1,2}\frac{|\mathcal{M}_{1+2\to X}|^2}{32\pi^3\,S}\int_m^\infty dE\, f_X^{\rm eq}\left(\beta p+2T\ln\left[\frac{1\mp e^{-E_+/T}}{1\mp e^{-E_-/T}}\right]\right), 
\end{align}
where $E_\pm=(E\pm\beta p)/2$, $\beta=\sqrt{1-4m_1^2/m^2}$, and $S$ is the symmetry factor. 
The yield mass density after reheating is
\begin{align}
\label{eq:yield}
Y_X(m,\tRH)=\int_{m/\TRH}^\infty \tilde g(x)\,\frac{R_{\rm ID}(m,x)}{x\,H(x)\,s(x)}\,dx,
\end{align}
with $x=m/T$, $H=(\pi^2 g_* T^4/90M_P^2)^{1/2}$, $s=2\pi^2 g_{*,s}T^3/45$, and $\tilde g(T)=1+\frac{1}{3}\,\dfrac{d\ln g_{*,s}}{d\ln T}$, as in the main text. Also, we use the notation where $R_{\rm ID}(m,x)$ is identified with $R_{\rm ID}(m,T)$ with the replacement $T\to m/x$. 

The total energy density contained in the tower today assuming approximate mass conservation is
\begin{align}
\rho_{X,0}=s_0\int_0^\infty m\,Y_X(m,\tRH)\,\frac{dm}{\mKK},
\end{align}
with $s_0\simeq2.2\times10^{-38}~\GeV^3$~\cite{Planck:2018vyg}, as in the main text. 

We assume the axion couples only to photons, $\mathcal{L}\supset-\frac{1}{4}\ga\,aF_{\mu\nu}\tilde F^{\mu\nu}$, with
\begin{align}
\label{eq:Msq_axion}
|\mathcal{M}_{\gamma\gamma\to a}|^2=\frac{1}{2}\,\ga^2\,m^4,
\end{align}
summed over the two photon polarizations and where $S=2$.

The graviton couples universally to the stress tensor, $\mathcal{L}\supset-h_{\mu\nu}T^{\mu\nu}/M_P$, and is produced from every bath pair. Summing over the five graviton polarizations and over the initial spins and colors~\cite{Han:1998sg},
\begin{align}
\label{eq:Msq_graviton}
|\mathcal{M}_{\gamma\gamma\to h}|^2&=\frac{2m^4}{M_P^2}\,,\qquad
|\mathcal{M}_{gg\to h}|^2=\frac{16\,m^4}{M_P^2}\,,\qquad |\mathcal{M}_{f\bar f\to h}|^2=\frac{N_c}{2}\left(1-4r_f\right)\left(1+\frac{8r_f}{3}\right)\frac{m^4}{M_P^2}\,,
\end{align}
with $r_f=m_f^2/m^2$, $N_c=1\,(3)$ for leptons (quarks), $S=2$ for the identical bosons $\gamma\gamma$ and $gg$, and $S=1$ otherwise. A Weyl neutrino counts as half a massless Dirac fermion, so the three neutrinos give $3m^4/M_P^2/4$.

The sum in $R_{\rm ID}$ runs over initial states in the thermal bath. The temperatures of greatest interest to dark dimension dark matter occur not too high above the QCD phase transition which occurs around $T_{\rm QCD}\approx 156$ MeV~\cite{HotQCD:2018pds}. We perform an approximation where for $T> T_{\rm QCD}$ MeV we add massless $u$, $d$, $s$, $c$ and $g$ to graviton inverse decay rates. For $T< T_{\rm QCD}$ MeV we only add pions. In both cases we also include all leptons and photons. Fortunately, the temperatures where graviton freeze-in production is of primary interest occur for $\TRH\gtrsim 500$~MeV, such that we are not extremely sensitive to this transition.

\subsection{Discussion on Robustness to Early Universe Cosmology}\label{app:EMD}

Given the UV dependence of the freeze-in calculation, it is important to understand how the calculation depends on times before reheating. For example, what if there is a period of early matter domination (EMD)?

During a long period of early matter domination, radiation energy density will be subdominant with a temperature which redshifts as $T\propto R^{-3/8}$, with $R$ being the scale factor. Assuming the production of KK excitations arises dominantly from freeze-in, the production rate density scales as $T^6/\Lambda_{\rm UV}^2$ (a per-particle rate $\Gamma \sim T^3/\Lambda_{\rm UV}^2$ acting on the $\propto T^3$ bath density), with $\Lambda_{\rm UV}$ a UV scale associated with the strength of the interaction. During an EMD era $\frac{d}{dt}  = \frac{dR}{dt} \frac{d}{dR} = H \frac{d}{d \log R} = -\frac{3}{8}H \frac{d}{d \log T}$. Then using $H \propto T^4$ gives
\begin{align}
    \frac{d n_k}{d\log T}\propto -T^{2}\implies n_k \propto T_{\rm max}^2\,.
\end{align}
The energy density at $T_{\rm max}$ (which could be much larger than the final reheating temperature, $T_{\rm RH}$) is obtained by multiplying by the number of modes and the mass (both $\propto T_{\rm max}$), giving
\begin{align}
    \rho_{KK}(T_{\rm max}) \propto T_{\rm max}^4\,.
\end{align}
The energy density produced at $T_{\rm max}$ redshifts during the EMD era and contributes to the energy density at $\TRH$ proportional to $T_{\rm max}^{-4}$; therefore, this abundance must be subdominant. Mathematically this is described by
\begin{align}
    \rho_{KK}(\TRH) = \rho_{KK}(T_{\rm max}) \times \left(\frac{R_{\rm RH}}{R_{\rm max}}\right)^{-3}
    =
    \rho_{KK}(T_{\rm max}) \times \left(\frac{T_{\rm max}}{\TRH}\right)^{-8}\propto T_{\rm max}^{-4}\,.
\end{align}
The contribution produced before final reheating is therefore negligible, so the surviving KK tower relic density is fixed by the final reheating temperature $\TRH$ alone, independently of the earlier thermal history.

\section{Exact Analytic Single-Tower Fragmentation} \label{app:Calculation of Tower Fragmentation}

This appendix describes analytic results for tower fragmentation before towers reach the attractor solution to show:
\begin{enumerate}
    \item Distributions converge to the attractor solution.
    \item Initial conditions get washed out leaving only the information of the total mass (assuming mass conservation).
    \item The time scale to reach the attractor solution is identified with the lifetime of initial-state particles.
\end{enumerate}
Consider uniform fragmentation of a single tower with $F(m,m') = F(m)$ for $m>m'$. We Laplace transform Eq.~\eqref{eq:y_m_t_equation}
\begin{align}
    s\widehat{Y}(m,s) - Y_0(m) = -\Gamma(m)\widehat{Y}(m,s) +\int_m^\infty dm'\, \Gamma(m')F(m')\widehat{Y}(m',s),
\end{align}
where $Y_0(m) = Y_0(m,t = 0)$. We then differentiate with respect to $m$ (denoted by a prime) to get
\begin{align}
    \left(s + \Gamma(m)\right)\widehat{Y}'(m,s) + \left(\Gamma'(m) + \Gamma(m)F(m)\right)\widehat{Y}(m,s) = Y'(m,0)\,.
\end{align} 
Define $P(m,s) = \frac{\Gamma' + \Gamma F}{s + \Gamma}$, $Q(m,s) = \frac{Y'_0}{s + \Gamma}$, and $I(m,s) = \exp\left(\int_0^m dm' P(m',s)\right)$ to get
\begin{align}
    \widehat{Y}(m,s) = \frac{-1}{I(m,s)}\int_m^\infty dm' I(m',s)Q(m',s)\,.
\end{align}
If we specify $\Gamma(m) = m^p/m_\star^{p-1}$, where $m_\star$ is a normalization scale and define $\mu = m/m_\star$ and $\tau = m_\star t$, then
\begin{align}
    I(m,s) = \left(\frac{s+\mu^p}{s}\right)^{\frac{p+2}{p}}\,.
\end{align}
First consider initial conditions $Y_0(\mu) = \delta(\mu - \mathcal{M}_0)$ to get 
\begin{align}
    \widehat{Y}(\mu,s) =\begin{cases}
        0\,,& \mu> \mathcal{M}_0\\
        \dfrac{2 \mathcal{M}_0^{p-1}}{(s+\mathcal{M}_0^p)(s+\mu^p)}\left(\dfrac{s+\mathcal{M}_0^p}{s+\mu^p}\right)^{\frac{2}{p}}\,,& \mu<\mathcal{M}_0
    \end{cases} 
\end{align}
The inverse Laplace transform gives the solution
\begin{align}
    Y(\mu,\tau;\mathcal{M}_0) =\begin{cases}
        0\,, & \mu>  \mathcal{M}_0\\
        \dfrac{2 \mathcal{M}_0^{p-1} e^{- \tau \left(\mu^p+\mathcal{M}_0^p\right)/2} \mathcal{M}_{-\frac{2}{p},\frac{1}{2}}\left(\left(\mathcal{M}_0^p-\mu^p\right)
   \tau\right)}{\mathcal{M}_0^p-\mu^p}\,,& \mu<\mathcal{M}_0
    \end{cases} 
\end{align}
where $\mathcal{M}$ is the Whittaker function. For $(\mathcal{M}_0^p - \mu^p) \tau \gg 1$ this approaches the attractor solution 
\begin{align}
    Y(\mu,\tau;\mathcal{M}_0) \approx \begin{cases}
        0\,, & \mu> \mathcal{M}_0\\
        \dfrac{ \mathcal{M}_0 p  }{\Gamma
   \left(\frac{2}{p}\right)}e^{-\mu^p \tau}  \tau^{\frac{2}{p}}\,,& \mu<\mathcal{M}_0
    \end{cases} 
\end{align}
This approximate solution respects mass conservation and is described by a generalized gamma distribution, 
\begin{align}
\frac{1}{\mathcal{M}_0}\mu Y(\mu,\tau) = f_\Gamma(\mu ; a,d,p)=\frac{\left(p / a^d\right) \mu^{d-1} e^{-(\mu / a)^p}}{\Gamma(d / p)}\,, \qquad {\rm with}~ a = \tau^{-1/p}~{\rm and}~d = 2\,.
\end{align}
Solutions for general $Y_0(\mu)$ are constructed from the above by linearity and simplify at late times
\begin{align}
\mu Y(\mu, \tau) \approx \mathcal{M}_0\,f_\Gamma(\mu ; \tau^{-\frac{1}{p}},2,p) \,,\quad {\rm for}~ \mathcal{M}_0 = \int_0^\infty \mu' Y_0(\mu') d\mu'\,.
\end{align}

\begin{figure}[t]
    \centering
    \includegraphics[width=1.\linewidth]{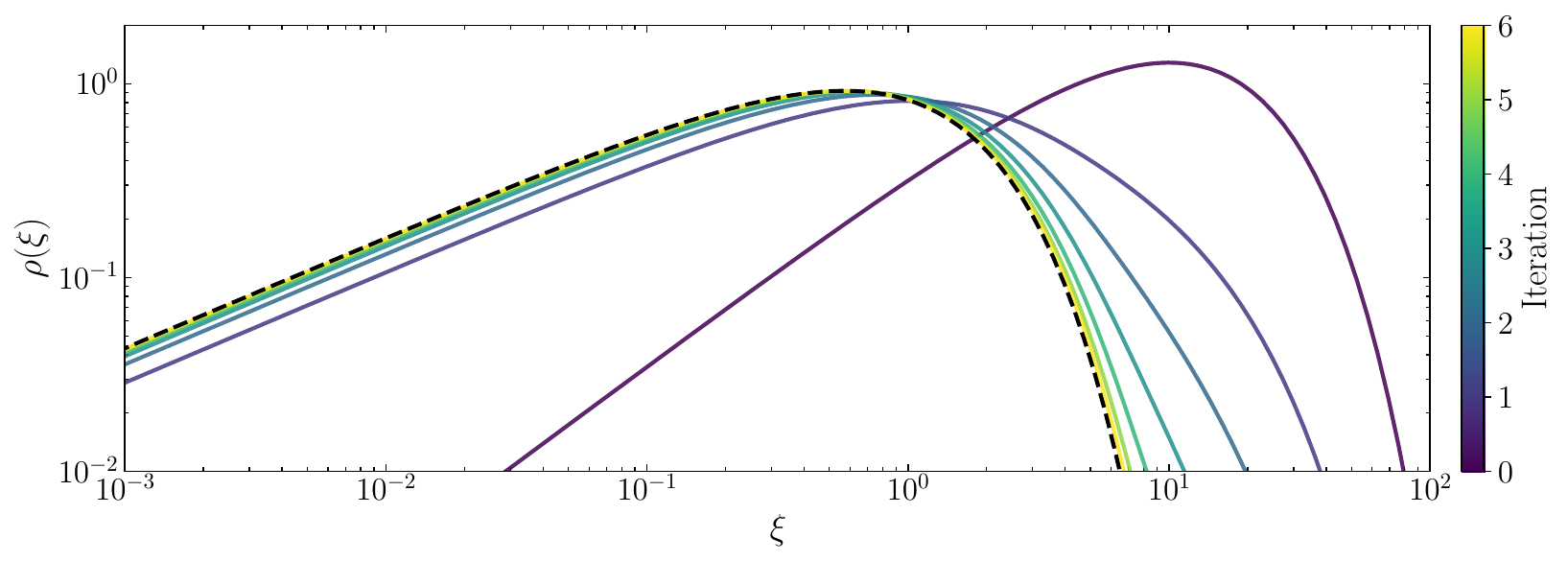}
    \caption{Iterative solution of Eq.~\eqref{eq:f_sol} for $g(x) = 2$, starting from a trial $\Psi^{(1)}\propto\xi^{2/p}e^{-\xi/10}$ (dark purple); each iteration is normalized to unit total mass, $\mathcal{M}=1$. By the sixth pass the profile is close to the exact attractor \Eq{eq:fdist} (black dashed).}
    \label{fig:iterations}
\end{figure}

\section{Numerical Results for Tower Fragmentation}\label{app:Numerical Fragmentation}

This appendix details the numerical solution of the coupled attractor system, Eqs.~\eqref{eq:diffeq_u_multi}--\eqref{eq:multi-tower_evolution}. We work with the uniform kernel $g(x)=2$, for which the single-tower attractor \Eq{eq:fdist} is known exactly and serves as a benchmark.

We treat all $N$ axion towers identically. The $N{+}1$ equations reduce to two profiles $(\Psi_h,\Psi_a)$. Writing $\mathcal{G}[\Psi]$ for the single-tower gain functional of \Eq{eq:G}, the components of the source \Eq{eq:source_multi} read
\begin{align}\label{eq:GhGa}
    G_h = \mathcal{G}[\Psi_h] + \frac{N\alpha}{2}\,\mathcal{G}[\Psi_a]\,,
    \qquad
    G_a = \alpha\,\mathcal{G}[\Psi_h] + \frac{\alpha}{2}\,\mathcal{G}[\Psi_a]\,.
\end{align}
We sample $\xi$ on a logarithmic grid of $n\simeq3000$ points spanning $\xi\in[10^{-8},100]$. 

The solution of the integral equation, \Eq{eq:multi-tower_evolution}, is advanced cell by cell. With $\delta_k=\xi_{k+1}-\xi_k$ and $y_k=\Lambda_i\,\delta_k$,
\begin{align}\label{eq:expstep}
    \Psi_i(\xi_{k+1}) = e^{-y_k}\,\Psi_i(\xi_k)
    + \int_{\xi_k}^{\xi_{k+1}} e^{-\Lambda_i(\xi_{k+1}-\xi')}\,G_i(\xi')\,d\xi'\,,
    \qquad
    y_k \equiv \Lambda_i\left(\xi_{k+1}-\xi_k\right)\,.
\end{align}
The complete algorithm, refining the iterative method of the main text, is:
\begin{enumerate}
    \item Initialize $\Psi_i^{(1)}(\xi)=\xi^{(s+2)/p}\,e^{-\Lambda_i\xi}$, with the small-$\xi$ exponent fixed by mass conservation.
    \item Evaluate $\mathcal{G}[\Psi_h^{(j)}]$ and $\mathcal{G}[\Psi_a^{(j)}]$ and assemble the sources \Eq{eq:GhGa}.
    \item Advance \Eq{eq:expstep} from $\Psi_i(\xi_{\rm min})=0$ to $\xi_{\rm max}$, taking the integral over each cell in closed form:
    with $\delta_k=\xi_{k+1}-\xi_k$ it equals $A_k\,G_i(\xi_k)+(A_k-B_k)\big[G_i(\xi_{k+1})-G_i(\xi_k)\big]$,
    where $A_k=(1-e^{-y_k})/\Lambda_i$ and $B_k=[1-(1+y_k)e^{-y_k}]/(\Lambda_i^2\delta_k)$,
    the latter evaluated by its series $\delta_k(\frac{1}{2}-\frac{y_k}{3}+\dots)$ for $y_k\lesssim10^{-4}$ to avoid cancellation.
    \item Normalize both profiles by the total mass $\mathcal{M}_{\rm tot} = \mathcal{M}_h+N\mathcal{M}_a$ from \Eq{eq:mass_conservation}.
    \item Repeat from step 2 until stationary.
\end{enumerate}
A few hundred passes leave the profiles stationary at the $10^{-4}$ level, and every number quoted below is unchanged for $n$ between $1500$ and $6000$.

Figure~\ref{fig:iterations} validates the method on a single tower ($\Lambda=1$), starting from a trial peaked an order of magnitude above the attractor. The sixth iterate is indistinguishable from it on the scale of the figure. The fixed point is independent of the trial profile, the numerical counterpart of the washout of initial conditions established analytically in App.~\ref{app:Calculation of Tower Fragmentation}.

Figure~\ref{fig:distributions} (left) shows the coupled solutions for $\alpha=1$ and $N=10,\,10^2,\,10^3$. The axion towers share the mass equally, $\mathcal{M}_a\approx\mathcal{M}_{\rm tot}/N$, while the graviton fraction falls as $\chi=0.152,\,0.047,\,0.013$, consistent with the scaling $\chi\propto N^{-(s+2)/p}=N^{-4/7}$ derived in the main text. The graviton peak slides to smaller masses as $\mu_h^{\rm peak}/\mu_a^{\rm peak}\propto N^{-1/p}$. The graviton profile is not, however, a rescaled copy of the single-tower attractor and above its peak, $\Lambda_h\xi\gg1$, the distribution is determined entirely from the axion distributions decays, i.e.
\begin{align}\label{eq:slaved}
    \Psi_h(\xi)\approx\frac{G_h(\xi)}{\Lambda_h}
    \xrightarrow{N\alpha\gg1 }
    \frac{1}{2}\,\mathcal{G}[\Psi_a](\xi)\,,
\end{align}
The high mass tail of the graviton distribution is what determines the rate of decays to the SM. Specifically, the energy injection rate per mass e-fold is proportional to $\mu^3 \Psi(\mu)$. The right panel of Fig.~\ref{fig:distributions} illustrates how the $\mu^3$ weighting fixes the primary source of graviton energy injection to the SM to be from modes with mass near the axion mass distribution peak.

Integrated and normalized to a single tower carrying the full DM mass, the graviton injection is suppressed by $0.39/N$, $0.43/N$, $0.45/N$ at $N=10,\,10^2,\,10^3$, approaching the asymptotic $\varepsilon_h(N)\approx0.46/N$ of \Eq{eq:eps}, and the same construction for a single visible axion tower gives $\approx0.69/N$. The bounds in the main text therefore reinterpret CMB bounds obtained from \cite{Law-Smith:2023czn} using the \emph{dominant peak mass} $m_0$, with the above abundance suppression factors scaling as $1/N$ for the graviton and for any visible axion tower.

\end{document}